\documentclass[9pt,conference]{IEEEtran}
\usepackage{amssymb,amsthm,amsmath,array}
\usepackage[mode=buildnew]{standalone}
\usepackage{graphicx}
\usepackage[caption=false,font=footnotesize]{subfig}
\usepackage{xspace}
\usepackage[sort&compress, numbers]{natbib}
\usepackage{stmaryrd}
\usepackage{xcolor}
\usepackage{mathtools}
\usepackage{float}
\usepackage{textcomp}
\usepackage{tikz}

\newcommand{\bs}{\boldsymbol}

\newcommand{\ie}{\emph{i.e.},\xspace}

\begin{document}
\title{Scan Coil Dynamics Simulation for Subsampled \\ Scanning Transmission Electron Microscopy}
\author{\IEEEauthorblockN{
        D. Nicholls\IEEEauthorrefmark{1}, 
        J. Wells\IEEEauthorrefmark{1}, 
        A. W. Robinson\IEEEauthorrefmark{1},
        A. Moshtaghpour\IEEEauthorrefmark{1}\IEEEauthorrefmark{2}, 
        A. I. Kirkland\IEEEauthorrefmark{2}\IEEEauthorrefmark{3}, and 
        N. D. Browning\IEEEauthorrefmark{1}
    }
    \IEEEauthorblockA{
        \IEEEauthorrefmark{1}University of Liverpool, UK, 
        \IEEEauthorrefmark{2}Rosalind Franklin Institute, UK,
        \IEEEauthorrefmark{3}University of Oxford, UK}
}
\maketitle

\begin{abstract}
    Subsampling and fast scanning in the scanning transmission electron microscope is problematic due to scan coil hysteresis - the mismatch between the actual and assumed location of the electron probe beam as a function of the history of the scan. Hysteresis limits the resolution of the microscope and can induce artefacts in our images, particularly during flyback. In this work, we aim to provide insights on the effects of hysteresis during image formation. To accomplish this, a simulation has been developed to model a scanning system as a damped double-harmonic oscillator, with the simulation being capable of managing many microscope dependant parameters to study the effect on the resultant scan trajectories. The model developed shows that the trajectory of the electron beam probe is not obvious and the relationship between scanning pattern and probe trajectory is complex.
\end{abstract}

\section{Introduction}
\begin{figure}[!t]
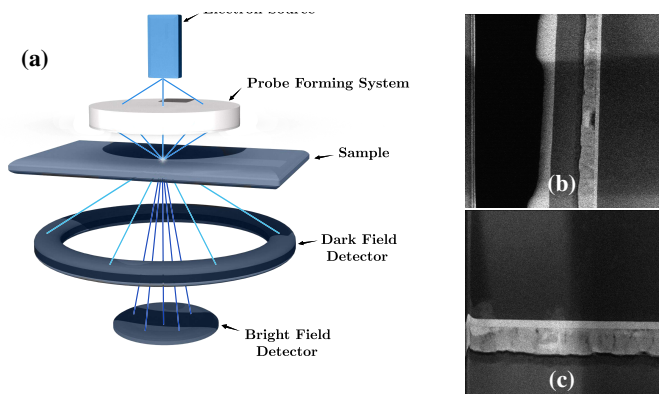

    \centering
    \includestandalone[width=1\columnwidth]{Figures/fig_stem_scheme}
    \caption{(a) Operating principles of STEM. Electrons are emitted from the source and subsequently condensed by the probe forming system. The probe is then raster scanned over the desired sampling area and the transmitted electrons are detected by either a dark field or bright field detector. The formed image is the intensity of the resulting electron wave-function at each pixel, corresponding to a certain set of scan coordinates. HAADF STEM images of a layered material oriented perpendicular (b) and parallel (c) to the fast scan direction, with no flyback time. A reflection of the image is present on the left hand edge, as a result of flyback distortions. STEM images courtesy of Dr. Mounib Bahri and Alex Robinson.}
    \label{fig:stem}
\end{figure}

Scanning transmission electron microscopy (STEM) \cite{crewe1974scanning, muller2009structure, pennycook2011scanning} utilises a focused electron beam probe, which is scanned across a field-of-view (FOV) of a sample using a scan coil, where a variety of signals are acquired at each position of the scan; see Fig.~\ref{fig:stem}. Examples of these collected signals are: \textit{(i)} unscattered or low-angle scattered electrons (as in bright field imaging), \textit{(ii)} high-angle scattered electrons (as in dark field imaging), and \textit{(iii)} x-rays (as in energy dispersive x-ray spectroscopy).
This is traditionally done using a raster scan - a space filling scan in which a FOV is sampled line by line, with the probe moving linearly across each line. This is typically performed left-to-right and top-to-bottom in STEM. 

When performing alternative sampling (\ie non-raster scanning) \cite{sang2016dynamic, velazco2022reducing} or subsampling, as in compressive sensing electron microscopy \cite{binev2012compressed, stevens2014potential, kovarik2016implementing, li2018compressed, nicholls2020minimising, nicholls2021subsampled, nicholls2022compressive, robinson2022sim, nicholls2022targeted, robinson2022sim2}, scan coil hysteresis becomes a major limiting factor. Scan coil hysteresis, often referred to simply as ``hysteresis'', is the disconnect between the assumed location of the scanning probe (as determined by the scanning pattern) and the actual location of the probe when moving from one scan location to another. This difference between the assumed and actual locations is due to the inductance on the electromagnetic scan coils, and is a problem as the STEM is constantly acquiring data, regardless of whether the probe has reached the scan co-ordinate or not. This can lead to the unintentional mapping of data to an inappropriate pixel, which presents itself as an image distortion or artefact. This effect scales with the distance traveled, such that large movements of the probe require longer settling times, and generally invoke more distortions.

In traditional raster scanning, the only time when the probe is not moving by only a small amount (small relative to the image size) is during flyback - when a line has finished, the probe must travel to the opposite edge to begin the next line again. This is problematic as it causes a distortion to manifest along the left-hand edge for every row. 
These flyback distortions have been accounted for in modern implementations of STEM by the addition of a flyback time, wherein the microscope holds the probe still after positioning the beam on the left hand edge for enough time such that time-dependent phenomena have passed and the beam remains stable. 

This method of sampling, using a raster scan with a flyback time, is appropriate for traditional application techniques. When attempting to perform fast scanning STEM \cite{buban2010high}, however, the removal of the flyback time is a natural method to increase acquisition times - doing so can remove between 100$\mu$s and 500$\mu$s per line (depending on microscope specifications), which at the extreme end equates to half a second per image (for an image with 1024 rows). This is significant, considering typical per-pixel acquisition times (dwell times) are on the order of microseconds. Methods of rectification for hysteresis distortions have shown positive results  \cite{jones2013identifying, mullarkey2022using}. Fig.~\ref{fig:stem} shows high-angle annular dark field (HAADF) STEM images of a focused ion beam prepared lamella oriented either perpendicular or parallel to the raster scan path with the flyback time set to zero. In Fig.~\ref{fig:stem}b, a reflection of the image is visible in the left hand edges of the images due to scan coil hysteresis. In Fig.~\ref{fig:stem}c, a down and uptick in the layer is witnessed, which is unexplained insofar.

\begin{figure*}[!t]
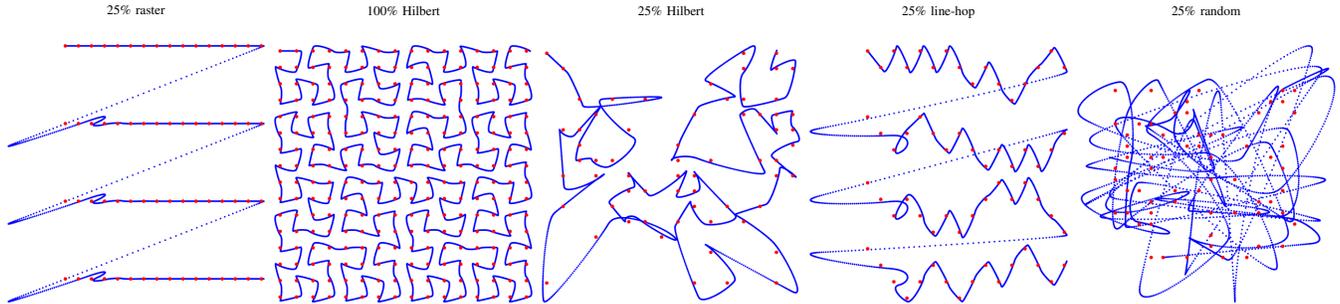

    \centering
    \includestandalone[width=1\textwidth]{Figures/fig_results}
    \caption{Example simulated scan trajectories for a set of scan co-ordinates spanning a 16x16 grid with a dwell time of 2 $\mu s$. Red dots are the scan co-ordinates associated with the labeled sampling pattern. Blue dots are locations of the probe during scanning at 1/32nd $\mu s$ intervals.} 
    \label{fig:example_scans}
\end{figure*}
Furthermore, the application of alternative and subsampled scanning to STEM is considerably complex, as not only are large movements of the probe often necessary using these techniques, the distance and angle between successive scan co-ordinates can change often, depending on the design of the scanning pattern. The image artefacts, such as the inability to perform uniform density sampling in STEM, that arise when employing these techniques are often difficult or impossible to explain beyond stating that hysteresis is the cause.

To better understand the effects of scan coil hysteresis, a model has been created to emulate the trajectory of the STEM probe during scanning at various parameter settings. A number of examples are presented here to provide supporting evidence as to why experiments regarding alternative scanning and subsampling in STEM are difficult to predict.

\section{Hysteresis Modelling \& Simulation}

 Proposed model assumes that the response of both \textit{(i)} scan coils and \textit{(ii)} electron beam are damped harmonic oscillators of the form:
\begin{equation}
    \frac{d^2\bs{x}}{dt^2} + 2\zeta \omega \frac{d\bs{x}}{dt} + \omega^2\bs{x} = 0,
\end{equation}
where $\bs{x}$ is the 2D displacement between the target location and the actual location, $t$ is time, $\zeta$ is the damping ratio of the oscillator, and $\omega$ is the undamped angular frequency of the oscillator. In this work, we consider different configurations of the damped harmonic oscillators model, while for the examples presented here, we set $\zeta_{\rm coil} = 1$, $\omega_{\rm coil} = 10$, $\zeta_{\rm probe} =  \sqrt{2}$, $\omega_{\rm probe} = \frac{1}{2 \sqrt{2}}$. Put simply, the coil response is fast, whilst the probe response is slow, relatively.

Fig.~\ref{fig:example_scans} shows five examples of scan trajectories following different sampling methods generated by this model: a 25\% raster scan (only every fourth row is sampled); a 100\% Hilbert scan (the scan follows a space filling Hilbert curve); a 25\% Hilbert scan (which follows the same curve, but with only a 25\% chance for a position to be included during the generation of the scan co-ordinates); a 25\% line hop curve, a commonly employed sampling method for subsampled STEM; and a 25\% random scan following a UDS scheme. 


The space filling Hilbert scan, whilst containing all scan co-ordinates within the scanned area, shows a distinct curve when moving between any two scan positions, but the pattern is relatively regular. The 25\% subsampled Hilbert scan is, however, unpredictably erratic in comparison. There is little coherence to the scan trajectory, and as such the predicted image generated from such a scan would be poor. The simulated line hop scan, which is emulating a popular option for practical subsampled STEM, shows a good amount of coincidence between the scan trajectory and the scan co-ordinates, except for during flyback, where this method suffers equally to raster scanning. The final case, the 25\% random scan, is very poor, and explains why experiments with regards to UDS schemes in STEM are difficult.

\begin{figure}
    \centering
        \begin{tabular}{c c}
           \fbox{\scalebox{1}[-1]{\includegraphics[angle=90,width=0.45\linewidth]{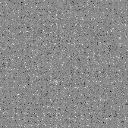}}} & \fbox{\scalebox{1}[-1]{\includegraphics[angle=90,width=0.45\linewidth]{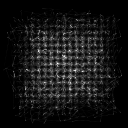}}} \\
           \fbox{\scalebox{1}[-1]{\includegraphics[angle=90,width=0.45\linewidth]{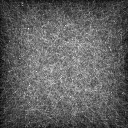}}} & \fbox{\scalebox{1}[-1]{\includegraphics[angle=90,width=0.45\linewidth]{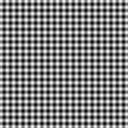}}}
        \end{tabular}
    \caption{(Top-left) Untreated simulated image produced by 100\% UDS. (Top-right) Image formed by assigning data to the actual probe locations. (Bottom-left) Density of scanned locations. (Bottom-right) Rectified image showing no distortions.}
    \label{fig:rectified}
\end{figure}

By assigning the pixel data to the actual position of the probe, rather than the assumed position (the scan co-ordinate), it should be possible to produce scan distortion free images. An example of this rectification method applied to a simulated approximation of a STEM image is shown in Fig.~\ref{fig:rectified}. The viability of a practical implementation of a similar rectification method is currently unknown, and warrants further investigation.

\section{Conclusion}
A simulation has been produced which models the scanning electron beam in a STEM as a damped double-harmonic oscillator system. This model qualitatively predicts hysteresis issues when scanning in STEM, by either traditional, alternative, or subsampled methods, and can be used to study the effects of certain parameters with regards to the scanning system and the sampling pattern generation. A rectification method is also proposed to address the issues of scan coil hysteresis. The future of this work will include quantitative validation of the model as well as using this model to help correct these distortions in experimentally acquired data. Further investigation into the efficacy of other sampling patterns is also planned, as well as potential application to other scanning instruments.

\vfill\pagebreak 

\bibliographystyle{IEEEtran}
\bibliography{references}
\end{document}